\pdfoutput=1
\documentclass[10pt,conference]{IEEEtran}
\usepackage{amsmath,amssymb,amsfonts}
\usepackage{color,soul}
\usepackage{mathtools}

\usepackage[export]{adjustbox}
\usepackage{caption}
\usepackage{url}
\usepackage{subfig}
\usepackage{algorithmic}
\usepackage{lettrine}
\usepackage{amsmath}
\usepackage{graphicx}
\usepackage{textcomp}
\usepackage[english]{babel}
\usepackage{amsthm}
\usepackage[ruled,vlined]{algorithm2e}

\theoremstyle{definition}

\DeclareGraphicsExtensions{.png,.pdf}
\usepackage{array}
\usepackage{multirow}
\usepackage{multicol,tabularx,capt-of}
\usepackage{multirow}
\usepackage{cite}
\usepackage{xcolor}
\usepackage{array,multirow,textcomp}
\theoremstyle{definition}

\theoremstyle{remark}
\usepackage{authblk}

\usepackage{float}
\usepackage{geometry} 
\usepackage{authblk}
\usepackage{etoolbox}
\IEEEoverridecommandlockouts\IEEEpubid{\makebox[\columnwidth]{  979-8-3503-1090-0/23/\$31.00~\copyright~2023 IEEE \hfill} \hspace{\columnsep}\makebox[\columnwidth]{ }}

\begin{document}
\title{Emergent Communication Protocol Learning for Task Offloading in Industrial Internet of Things}
\author[$*$]{Salwa Mostafa}
\author[$+$]{Mateus P. Mota}
\author[$+$]{Alvaro Valcarce}
\author[$*$]{Mehdi~Bennis}
\affil[$*$]{Centre for Wireless Communications, University of Oulu, FI-90014 Oulu, Finland.}
\affil[$+$]{Nokia Bell Labs, Nozay, France}
\affil[$*$]{salwa.mostafa, mehdi.bennis@oulu.fi}
\affil[$+$]{mateus.pontes\_mota@nokia.com, alvaro.valcarce\_rial@nokia-bell-labs.com}

\maketitle

\begin{abstract}

In this paper, we leverage a multi-agent reinforcement learning (MARL) framework to jointly learn a computation offloading decision and multichannel access policy with corresponding signaling. Specifically, the base station and industrial Internet of Things mobile devices are reinforcement learning agents that need to cooperate to execute their computation tasks within a deadline constraint. We adopt an emergent communication protocol learning framework to solve this problem. The numerical results illustrate the effectiveness of emergent communication in improving the channel access success rate and the number of successfully computed tasks compared to contention-based, contention-free, and no-communication approaches. Moreover, the proposed task offloading policy outperforms remote and local computation baselines. 
\end{abstract}

\begin{IEEEkeywords}
Dec-POMDP, IIoT, mobile edge computing, multiagent reinforcement learning, emergent communication.
\end{IEEEkeywords}

\section{Introduction}

The rapid development of 5G and beyond network technologies is enabling a plethora of Internet of Things (IoT) applications. The number of connected IoT devices is expected to reach 25.4 billion devices by~$2030$~\cite{WinNT}. The application of IoT in the industrial sector referred to as Industrial Internet of Things (IIoT) accounts for a large part of these connected devices~\cite{qiu2020edge}. In industrial environments, sensors, actuators, and other machines continuously generate various types
of computation tasks, leading to a large amount of data traffic, that needs to be processed in a timely, reliable, and efficient way. However, IIoT mobile devices have limited power and computation resources, which limit their ability to process their computation tasks in a reliable and efficient manner~\cite{khalil2021deep}. To tackle this problem, mobile edge computing (MEC) is a promising solution in which base stations are equipped with high-capacity computation resources, to which IIoT mobile devices can offload their computation tasks for fast execution. Nevertheless, computation task offloading adds an additional transmission delay to the computation delay. Therefore, IIoT mobile devices must decide whether to remotely or locally execute their computation tasks based on the availability and quality of communication resources and traffic load. 

The huge number of connected IIoT mobile devices and their various computation data traffic load drives the adaptability of dynamic multichannel access schemes to allow efficient utilization of spectrum resources. However, it is difficult for IIoT mobile devices to observe all channel states across the network. Traditional multichannel access approaches are mainly categorized into contention-based and contention-free schemes. In contention-based, IIoT mobile devices access the channel in a random way, thus if multiple IIoT mobile devices simultaneously choose the same channel to offload their computation tasks, high interference and collision may occur leading to low efficient task offloading performance~\cite{bockelmann2018towards,khalil2021deep}. On the other hand, contention-free adopts a centralized coordinator to allocate resources, which may cause overhead and delay, that may not be tolerable by some applications. Thus, task offloading may not give better performance than local computing. Therefore, in this work, we investigate the problem of joint task offloading decision and task scheduling for maximizing the number of computation tasks that can be executed under a task deadline constraint. Recently, reinforcement learning has shown a great advantage in solving problems in dynamic environments, where agents can make decisions based on their partial observations and historical information~\cite{feriani2021single}. Therefore, we adopt reinforcement learning (RL) as a framework to coordinate multichannel access and task offloading to achieve efficient computing. 

The works in~\cite{deng2021intelligent,chen2020deep,ren2020deep,hossain2020edge,cao2020multiagent} studied the task offloading problem in the IIoT environment using RL. Studies  in~\cite{deng2021intelligent,chen2020deep} minimized the system delay using different RL algorithms such as Q-learning, and deep deterministic policy gradient (DDPG). However, the authors ignored the multichannel access problem and assumed that communication channels are uniformly distributed and pre-allocated to the IIoT mobile devices at each time slot, which is an unrealistic assumption. The work in~\cite{ren2020deep,hossain2020edge} considered multiple edge computing servers to minimize the long-term energy consumption and system costs (i.e., energy and delay). Nonetheless, the authors assumed that each edge server has one communication channel serving an IIoT mobile device, which is an unrealistic assumption.

The closest related work to our study is~\cite{cao2020multiagent}, which considered the joint multichannel access and task offloading problem. Therein, the proposed RL algorithm shows a significant reduction in computation delay and improvement in channel access success rate compared to single-agent RL algorithms. However, the authors ignore the connection establishment time, during which the signaling control messages are exchanged among agents to coordinate channel access which may not always achieve a good policy. Moreover, the authors ignored dynamically generated computation traffic. In contrast to this work, we consider the problem of jointly learning the signaling control messages and offloading decision policy to achieve a better coordination policy in dynamic environments. 

Emergent communication protocols have been studied using MARL in~\cite{foerster2016learning,sukhbaatar2016learning} and have shown a significant improvement in learning cooperative behavior in multi-agent reinforcement learning. The authors in~\cite{valcarce2021toward,mota2021emergence,mota2022scalable,miuccio2022learning} show that a significant reduction in signaling overhead and delay and higher average good-put and successful channel accessing is achieved via emergent communication protocols. Thus, in this study, we aim to propose a general framework adopting emergent communication protocols for solving the computation offloading decision and multichannel access problem. The goal is to let agents learn a joint offloading decision and multichannel access policy via communication. To the best of our knowledge, this is the first work applying emergent communication protocols in a mobile edge computing IIoT scenario. Our contribution can be summarized as follows: 

\begin{itemize}
\item We proposed a novel framework for mobile edge computing in IIoT based on emergent communication to solve the problem of joint task offloading decision and scheduling of computation tasks.

\item We consider a dynamic traffic arrival model and show that our proposed combined scheme, which offloads part of the tasks when resources are available and executes part of tasks locally in case of scarce resources outperforms remote and local computation schemes in increasing the number of successfully computed tasks within the deadline constraints.

\item Simulation results demonstrate the efficiency of emergent communication in increasing the channel access success rate and the number of successful computation tasks within deadline constraints compared to traditional schemes and no communication approaches.
\end{itemize}

The rest of the paper is organized as follows. In Section~\ref{system}, we state our system model. In Section~\ref{Problem}, we formulate the problem using a reinforcement learning framework. Section~\ref{simulation} provides our simulation model and results. Finally, we conclude the paper in Section~\ref{conclusion}.

\section{System Model}\label{system}

\subsection{Network Model}

\begin{figure} 
   \centering
   \includegraphics[width= 3 in, height = 2.3 in]{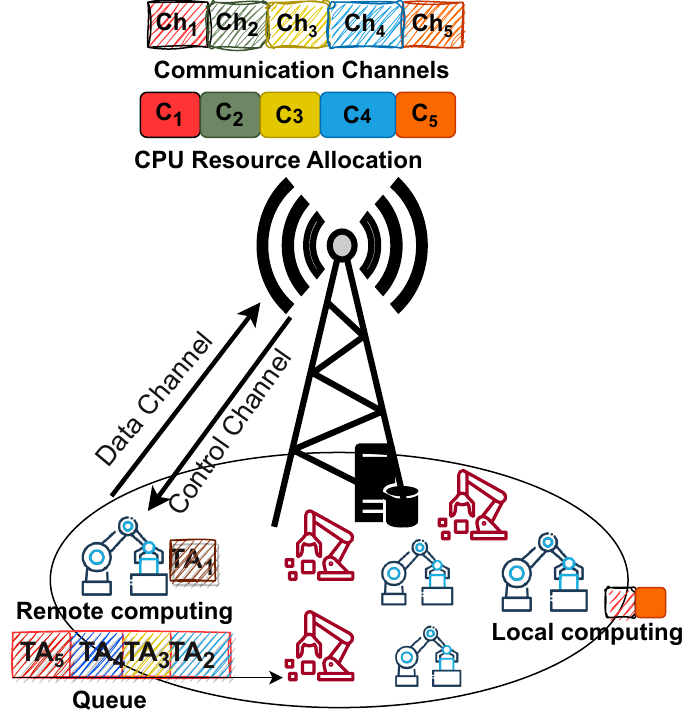}
    \caption{System Model.}
   \label{systemmodel}
\end{figure}

We consider a mobile edge computing (MEC) network system consisting of a base station and~$N$ industrial internet of things (IIoT) mobile devices indexed by~$\mathcal{N} \triangleq \{1,2,\dots,N\}$ as shown in Fig.\ref{systemmodel}. The base station has a central processing unit (CPU) with maximum computation speed~$F_{\mathrm{max}}$ and $M$ downlink multiaccess channels indexed by~$\mathcal{M} \triangleq \{1,2,\dots,M\}$ each with bandwidth~$W$~MHz. The base station computation resources are divided equally among the scheduled users to allow parallel computation. Each IIoT mobile device has limited computation speed~$f_n$ and a computation queue operates in a first-in-first-out (FIFO) manner with a maximum capacity of~$K$ computation tasks indexed by~$\mathcal{K}=\{1,2,\dots,K\}$. Each computation task~$k$ is non-dividable and has parameters~$(A_k,L_k,\tau_k)$, where~$A_k$ is the task size in bits,~$L_k$ is the number of CPU cycles per bit required for complete task execution and~$\tau_k$ is the task delay deadline~{\footnote{Note that we consider a smart logistics use case, where mobile robots read the RFID of goods and get information about the goods like size, type etc. then offload the information to the base station to compute the best path for goods from a vehicle to shelves and vice versa.}}. The arrival of computation tasks in the queue is modeled as a Poisson process with arrival rate~$\lambda = p_k \times T$, where~$p_k$ is the task arrival probability and~$T$ is the communication time period. The computation tasks can be either computed locally at the IIoT mobile device or remotely at the base station. The computation decision is denoted by~$x_{k,n} = \{0,1\}$, where~$x_{k,n}$ takes value $0$ if task~$k$ at mobile device~$n$ is locally computed and~$1$ if the task is remotely computed. We assume that each IIoT mobile device offloads the task with a fixed power level~$p_n$.

\subsection{Communication Model}

We assume that the base station and IIoT mobile devices communicate over data and control channels. The data channels are used for exchanging data while control channels are used to share information about the system state. We consider a discrete system model which adopts an orthogonal frequency division multiple access (OFDMA) transmission scheme.  At each time slot, the IIoT mobile devices randomly access the shared data channels to offload their computation tasks, which may lead to collisions. Thus, at each time step~$t$, the base station can transmit a control message~$D_n$ to each IIoT mobile device and each IIoT mobile device~$n$ can send a control message~$U_n$ to the base station while being able to offload his computation task through one of the uplink shared data channels to manage the channels access. We assume that the uplink and downlink control channels are dedicated and error-free and have a maximum capacity of~$C_{\mathrm{max}}$ bits per second, which allows transmission of messages from a vocabulary set~$\mathcal{V}$ of size~$2^{C_{\mathrm{max}}}$(i.e. each message has length $C_{\mathrm{max}}$ bits). We also assume that the data channels are modeled as a time-invariant Gaussian channel with additive white Gaussian noise vector $\boldsymbol{z} \sim\mathcal{CN}(0,\sigma^2)$, where $\sigma^2 = \aleph_0 W$ and $\aleph_0/2$ is the noise power spectral density. Thus, the uplink rate of IIoT mobile device~$n$ is expressed as 
\begin{equation}
R_n = W \log_{2}\left(1 + \frac{g_{n,m} p_n}{\sigma^2}\right)   \; \; \textit{bps}
\end{equation}
where~$g_{n,m}$ is the data channel gain between the base station channel~$m$ and IIoT mobile device~$n$. We assume that each IIoT mobile device can be allocated at most one channel at each time slot and offload one task. We consider the transmission of computation tasks to be finished when each IIoT mobile device's queue is empty. 

\subsection{Computation Model}

At the beginning of each time slot, each IIoT mobile device has a new computation task that arrives with a certain probability. We adopt the computation model based on the advanced dynamic voltage and frequency scaling (DVFS) technique at both the base station and IIoT mobile devices~\cite{you2016energy}. If the IIoT mobile device~$n$ decided to compute task~$k$ locally, the local computation time is given by
\begin{equation}
t^l_{k,n} = \frac{A_{k,n} \times L_{k,n}}{f_n}  
\end{equation}
If the IIoT mobile device~$n$ decided to offload task~$k$ to the base station, the remote computation time includes the upload time, execution time at the base station, and download time for the computed result. Since in our usecase (i.e smart logistics) the size of the computed result is relatively small compared to the task size and the base station has high power capabilities, we ignore the download time. Therefore, the remote computation time includes only the upload time and execution time at the base station, which is given by
\begin{equation}
t^r_{k,n} = t^u_{k,n} + t^{\mathrm{e}}_{k,n} =  \frac{A_{k,n}}{R_n} + \frac{A_{k,n} \times L_{k,n}}{f_m} 
\end{equation}
where~$f_m$ is the computation resources allocated to the computation task offloaded on channel~$m$. Hence, for IIoT mobile device~$n$, the delay time for task~$k$ can be expressed as follows 
$$t_{k,n} = \begin{cases}
t^l_{k,n} & \text{ if } x_{k,n} = 0 \; \; \text{local computation}\\
t^r_{k,n} & \text{ if } x_{k,n} = 1 \; \; \text{remote computation}
\end{cases}$$
In this study, our objective is to maximize the number of computation tasks that can be executed within the delay constraint by solving the problem of joint offloading decision and scheduling of computation tasks. Since successful channels access have a major impact on the task offloading decision, we also aim to maximize the channel access success rate expressed as~$R_s = \frac{N_s}{M \times T}$ which is defined as the total number of channels successfully accessed denoted by~$N_s$ divided by the total number of channels and the number of communication time slots. 

\section{Problem Formulation}~\label{Problem}

\begin{figure} 
   \centering
   \includegraphics[width= 3.2 in, height = 1.4 in]{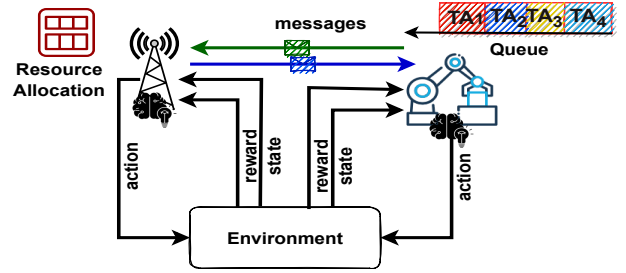}
    \caption{IIoT MDs and BS are cooperative MARL.}
   \label{MARL}
\end{figure}

We formulate the problem as a multi-agent reinforcement learning (MARL) cooperative task due to the advancements in deep reinforcement learning (DRL), where the base station and IIoT mobile devices are RL agents that need to learn how to communicate to solve the problem of joint offloading decision and scheduling of computation tasks as shown in Fig.~\ref{MARL}. The problem is modeled as a decentralized partially observable Markov decision process (Dec-POMDP), augmented
with communication. A Dec-POMDP for~$n$ agents is defined
by the global state space~$\mathcal{S}$, an action space~$\mathcal{A}_1,\mathcal{A}_2,\dots,\mathcal{A}_n$,
and an observation space~$\mathcal{O}_1,\mathcal{O}_2,\dots,\mathcal{O}_n$ for each agent. In Dec-POMDP, an agent observation does not fully describe the environment state and each agent action space is subdivided into an environment action space and a communication action space.

We assume that the agents interact with each other over $T$ finite episodes, where each episode terminates when the queues of all IIoT mobile devices are empty or the maximum number of time steps~$t_{\mathrm{max}}$ is reached. At each time step~$t$, each agent has his own observation of the system. The IIoT mobile device~$n$ has observation~$o^n_t = (|\mathcal{K}|,S)$, which includes the number of computation tasks in the queue~$|\mathcal{K}|$, and $S$ the selected channel state. The selected channel state takes values~$S \in \{0,1,2\}$ where~$0$ means that the channel is not needed in case of local computation,~$1$ if the channel is free and~$2$ if a collision occurs over the channel. The base station has an observation~$o^b_t = (h_1,h_2,\dots,h_M,\mathcal{N}_s)$, which contains the states of the $M$ channels, where
$$h_m = \begin{cases}
0, & \text{if the channel is idle}\\
n, & \text{if IIoT MD~$n$ transmits on the channel}\\
N + 1, & \text{if the channel has collision} 
\end{cases}$$
indicates if the channel is idle or has a successful transmission from user~$n$ (i.e. one IIoT mobile device transmits) or has a collision (i.e. more than one IIoT mobile device transmits). It also contains the mobile devices that successfully computed their tasks within the deadline constraint denoted by~$\mathcal{N}_s$. 

The action space contains the actions of the IIoT mobile devices and base station. Each IIoT mobile device based on its current state takes an environment action and a communication action. The environment action contains an offloading decision and a channel selection action~$\boldsymbol{a}^t_n = (a_o,a_c)$, where~$a_o \in \{0,1\}$ is the offloading decision action and $a_c \in \{0,1,2,\dots,M\}$ is the channel selection action such that~$0$ indicates that the IIoT mobile device does not need to access the channel if the task is computed locally and~$\{1,2,\dots,M\}$ is the selected channel for offloading. The communication action (i.e uplink message)~$U^n \in \{0,1\}$ is interpreted by the base station as~$0$ for a null message and $1$ for a scheduling request. The base station based on its current state utters a communication action to each IIoT mobile device~$D_n \in \{0,1,2,M,M+1\}$ that has no direct effect on the environment, which is interpreted as~$0$ for a null message,~$\{1,2,\dots,M\}$ for scheduling grant to one of the communication channels and~$M+1$ for acknowledgment of the successful computation of the offloaded task. Note that the agents have no prior knowledge about the meaning of the communication messages and they learn to associate meaning during training.

The reward at each time step is defined as
$$r_n(t) = \begin{cases}
+\rho & \text{ if the task computed within the deadline}\\ 
      & \text{constraint.}\\
-\rho & \text{if the task computation exceeded the} \\ 
      & \text{deadline constraint.} \\
0 & \text{ otherwise}
\end{cases}$$
The reward is $+\rho$ if the IIoT mobile device computed his task within the deadline constraint,~$-\rho$ if the task computation exceeds the deadline constraint, and~$0$ otherwise. The team reward is the sum of the rewards of all IIoT mobile devices, which is defined as~$r(t) = \sum_{n \in \mathcal{N}} r_n(t)$. The agent state at time step~$t$ is a tuple comprising the most recent~$l$ observations, actions, and uplink and downlink messages:
\begin{itemize}
\item IIoT mobile device state: $(o^n_t,\dots,o^n_{t-l},a^n_t,\dots, a^n_{t-l},U^n_t,\\\dots,U^n_{t-l},D^n_t,\dots,D^n_{t-l})$
\item Base station state: $(o^b_t,\dots,o^b_{t-l},\boldsymbol{U}_t,\dots,\boldsymbol{U}_{t-l},\boldsymbol{D}_t,\\\dots,\boldsymbol{D}_{t-l})$
\end{itemize}
where~$\boldsymbol{U} \triangleq [U_1,U_2,\dots,U_N]$ and~$\boldsymbol{D} \triangleq [D_1,D_2,\dots,D_N]$ are the uplink and downlink messages, respectively. Each IIoT mobile device deletes the computation task from the computation queue if it is successfully computed locally or remotely at the base station otherwise the task is kept in the queue for two-time steps and then dropped.

IIoT MDs have low computational power and low battery power which makes performing AI learning models at each device not practical and limits cooperation among MDs. Thus, we use the multi-agent proximal policy optimization (MAPPO) algorithm~\cite{schulman2017proximal,yu2021surprising}, which is an on-policy policy gradient algorithm. MAPPO allows centralized training and decentralized execution (CTDE), where the agents learn a shared optimal policy instead of individual policy for each agent. The actor-critic network architecture is adopted along with the generalized advantage estimation (GAE)~\cite{schulman2015high}. MAPPO optimizes a surrogate-clipped objective function and updates the policy and value functions via mini-batches.

\section{Simulation Model and Results}\label{simulation}

\begin{table}
\caption{Simulation Parameters}
\centering
\begin{tabular}{|c|c|}
\hline
\textbf{Parameters} & \textbf{Values}\\
\hline
Number of Sub-carriers  & $2$ \\
\hline
5G-NR frequency band (FR1)& $410 – 7125$~MHz\\
\hline
Sub-carrier Bandwidth & $10$~MHz\\
\hline
Number of IIoT Mobile Devices & $3$ \\
\hline
Distance-dependent Path loss & $128.1 + 37.6 \log_{10} d,$ dB\\
\hline
Tasks Size  & $100-500$ bits\\
\hline
Tasks Computation Requirement  & $10^2- 2 \times 10^4$\\
\hline
Tasks Delay Tolerance &$1 - 5$~millisecond\\
\hline
Noise Power Spectral Density & -174 dBm/Hz\\
\hline
Base station Computation Capacity & $100$~GHz  \\
\hline
IIoT Mobile Device Computation Capacity & $1$~GHz  \\
\hline
IIoT Mobile Device Queue Capacity & $25$\\
\hline
Probability of Task Arrival & $0.90$\\
\hline
Duration of episode & 25\\
\hline
\end{tabular}
\label{table:1}
\end{table}

In this section, we evaluate the performance of our proposed framework. We consider a $10 \times 10$~m$^2$ warehousing logistic area with a base station and~$N = 3$ IIoT mobile devices~\footnote{limited computing resources restricted us to small setting but the idea is applicable to large setting scenarios.}. The path loss between the base station and IIoT mobile devices is modeled as~$128.1 + 37.6 \log_{10}(d)$, where~$d$ is the distance in Kilometers. The system parameters are listed in Table~\ref{table:1}. MAPPO is implemented with the hyperparameters listed in Table~\ref{table:2}, where the policy and value functions are represented by separate MLP fully connected linear neural networks and optimized by Adam optimizer~\cite{kingma2014adam}.

\begin{table}
\caption{MAPPO Hyperparameters}
\centering
\begin{tabular}{|c|c|c|c|}
\hline
$\textbf{Hyperparameter}$ & $\textbf{Values}$ & $\textbf{Hyperparameter}$ & $\textbf{Values}$\\
\hline
Number of episodes & 10000 & Learning rate & $10^{-3}$ \\
\hline
Minibatch size & $128$ & Discount factor~$(\gamma)$& $0.99$\\
\hline
GAE parameter~$(\lambda)$ & $0.95$ & Clipping parameter~$(\epsilon)$ & $0.2$\\
\hline
VF coeff.~$(c1)$ & $0.2$ & Entropy coeff.~$(c2)$ & $0.2$ \\
\hline
Optimizer & Adam & Optimizer epsilon & $10^{-5}$\\
\hline
\end{tabular}
\label{table:2}
\end{table}

We evaluate the performance over the training process, where the solid lines represent the average performance in the evaluation episodes during the training and the shaded regions show the~$95\%$ confidence interval (CI). We compare the proposed combined mode (i.e. local and remote computation) solution with the following baselines 
\begin{enumerate}
\item \textbf{Local computation:} all IIoT MDs locally compute their computation tasks.
\item \textbf{Remote computation with no-communication:} all IIoT MDs remotely compute their computation tasks at the BS without exchanging messages.
\item \textbf{Remote computation with communication:} all IIoT MDs remotely compute their computation tasks at the BS and exchange messages.
\item \textbf{Contention-free:} the BS controls and schedules the transmission over the downlink data channels. Each IIoT MD sends a scheduling request if its computation queue is not empty and offloads the task if the BS sends a scheduling grant. The IIoT MD deletes the task from the queue if it is computed successfully within the deadline constraint and an ACK is received from the BS. At each time step, the BS assigns the available channels to the IIoT MDs sent scheduling requests. If the IIoT MD made a successful task computation simultaneously and sent a scheduling request, the BS will send an ACK and ignore the scheduling request.
\item \textbf{Contention-based:} Each IIoT MD transmits with a certain probability~$p_t$ if the computation queue is not empty and randomly accesses the channels.
\end{enumerate}

\begin{figure}
    \centering
    \includegraphics[width = 3.5 in, height = 2.6in]{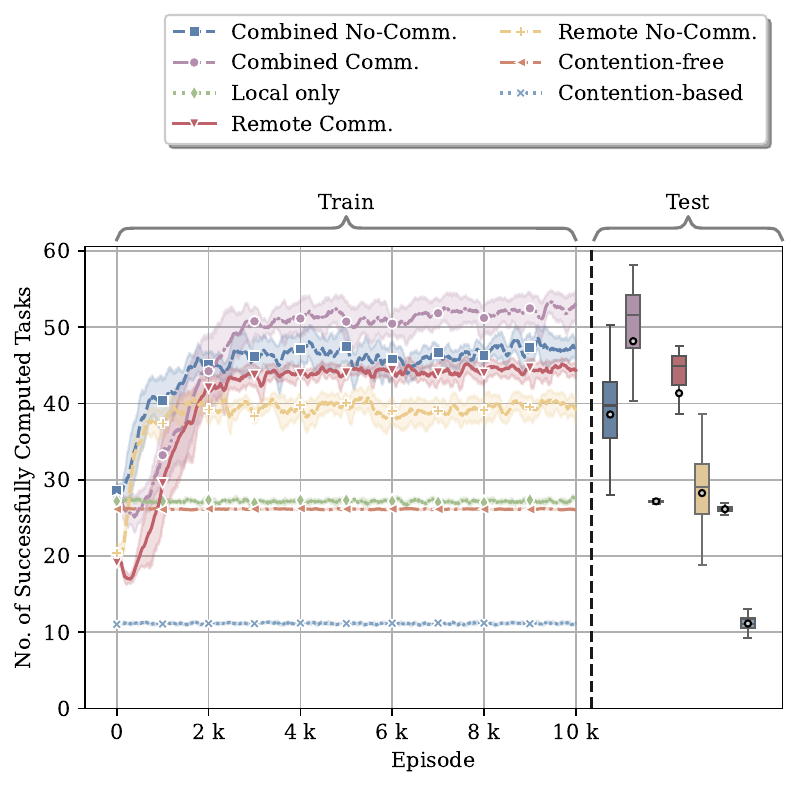}
    \caption{Number of tasks successfully computed versus training episodes.}
   \label{Fig1}
\end{figure}

Fig.~\ref{Fig1} shows the number of tasks successfully computed within the deadline constraint during training. It indicates the superiority of the proposed combined scheme with communication in increasing the number of successfully computed tasks within deadline constraints compared to other schemes during the training and testing phase as well. As we can observe, the combined and remote schemes with communication outperform the combined and remote with no communication highlighting the effectiveness of the learned communication protocol in improving system performance. Obviously, contention-free scheme outperforms the contention-based one as the BS schedules the transmission and eliminates collision compared to random channel access in contention-based. 

\begin{figure}
\centering
\includegraphics[width = 3.5 in, height = 2.6 in]{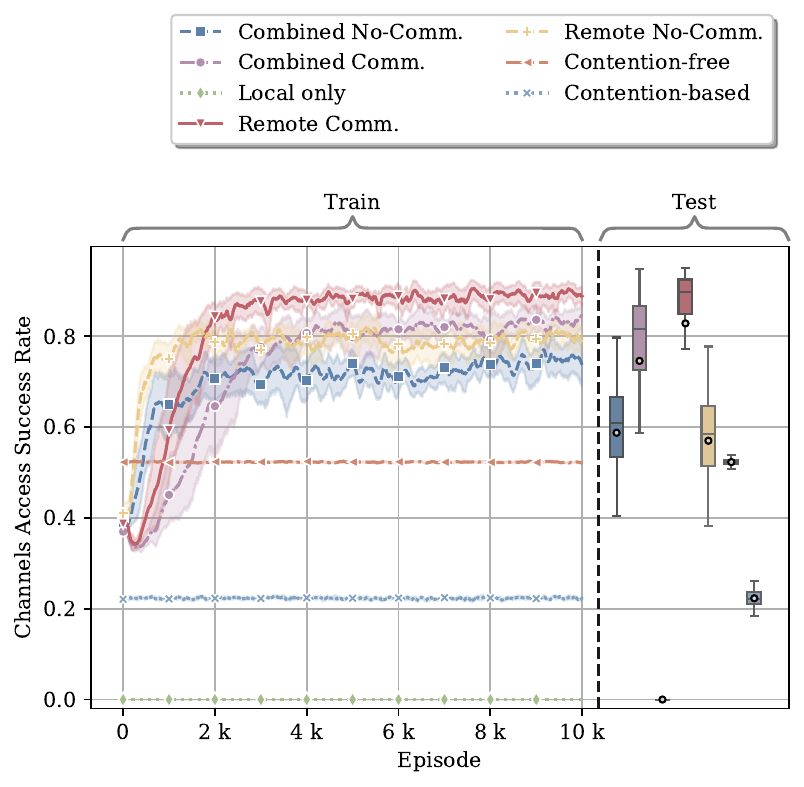}
\caption{Channel access success rate versus training episodes.}
\label{Fig3}
\end{figure}

 \begin{figure}
    \centering
    \includegraphics[width = 3.5 in, height = 2.6 in]{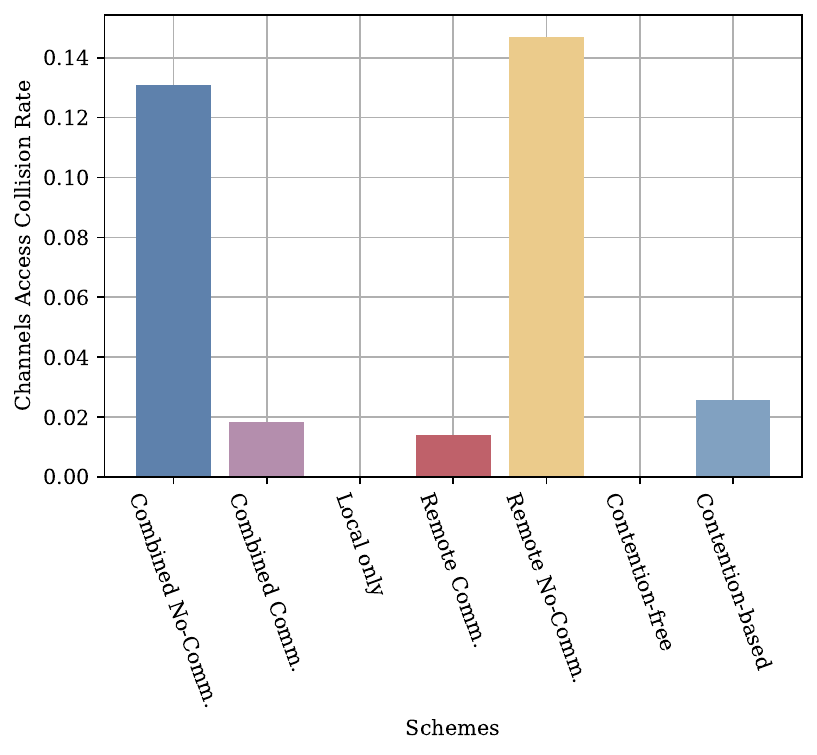}
    \caption{Collision rate for different schemes.}
   \label{Fig6}
\end{figure}

Fig.~\ref{Fig3} demonstrates the channels access success rate with training. As expected the remote scheme has the highest channel access rate as the IIoT MDs always offload their computation tasks to the BS. On the other hand, the local scheme does all computations locally and does not access the channels. As we can notice, the learned channel access protocol by remote schemes outperforms contention-free and contention-based schemes, which shows that the remote schemes learned different and more efficient protocols to access the channels. Moreover, the remote scheme with communication has a high channel access rate during the training and testing phases, which indicates that the IIoT MDs and BS learned a good communication protocol from scratch to coordinate channel access and reduce collision. The collision rate for different schemes is shown in Fig.~\ref{Fig6}. We can observe that the contention-free scheme has zero collision rate due to the centralized coordination of channels access. We can also observe the effectiveness of communication in remote and combined schemes in maintaining a low collision rate (i.e. high task offloading efficiency) compared to no communication schemes.

\begin{figure}
    \centering
    \includegraphics[width = 3.5 in, height = 2.6 in]{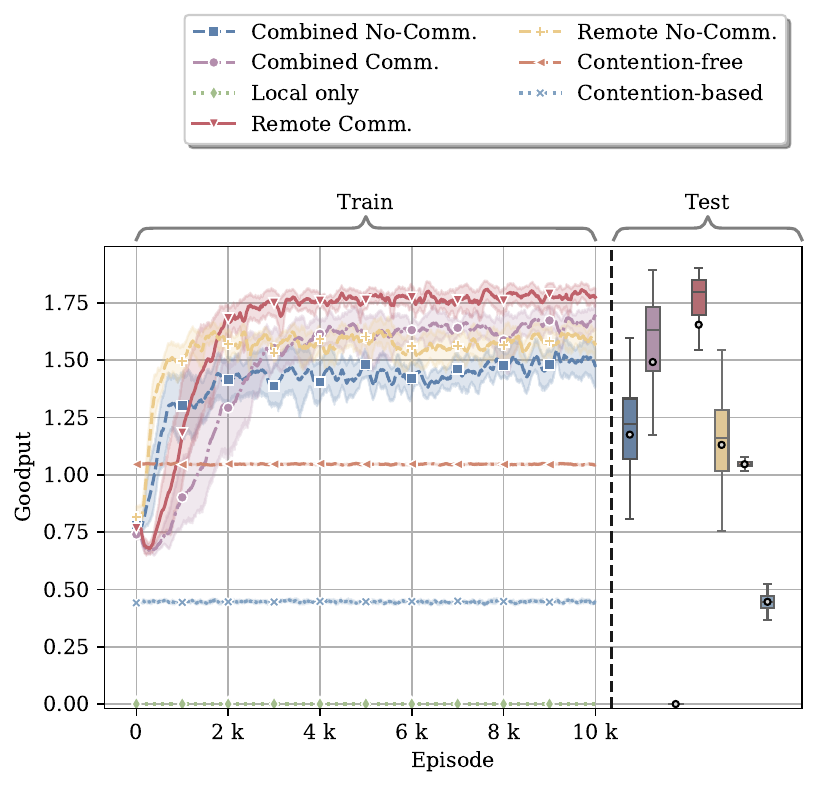}
    \caption{Goodput versus training episodes.}
   \label{Fig4}
\end{figure}

Fig.~\ref{Fig4} depicts the goodput versus the training episodes. The goodput is defined as the unique number of computation tasks successfully received at the base station divided by the episode duration. The larger the goodput the better the scheme performance as a larger number of tasks has been successfully received in less time. It is easy to see the superior performance of the remote scheme as it always sends computation tasks to the BS.

\section{Conclusion}\label{conclusion}

In this article, we proposed an emergent communication protocol learning framework for solving the problem of joint task offloading decision and scheduling of computation tasks in an IIoT scenario. The problem is formulated to maximize the number of computation tasks that can be executed within the deadline constraint. MARL framework is adopted where the base station and IIoT MDs are reinforcement learning agents that learn how to communicate with each other to solve the problem in a cooperative manner. The simulation results indicated the effectiveness of the learned protocols in maintaining highly efficient task offloading and maximizing the number of successfully computed tasks within the deadline constraint compared to traditional approaches. In future work, the scalability of the proposed approach in which a large number of IIoT mobile devices are connected to the network will be studied. 

\section{ACKNOWLEDGMENT}

The work is funded by the European Union through the projects CENTRIC (G.A no. 101096379), and VERGE (G.A no. 101096034) and the European Commission through the projects HORIZON SNS JU ADROIT6G (G.A no. 101095363) and  DESIRE6G (G.A no. 101096466). The work is also supported by the project SCENE (G.A no. 00164501.0). Views and opinions expressed are however those of the author(s) only and do not necessarily reflect those of the European Union. Neither the European Union nor the granting authority can be held responsible for them.  We would like to thank Nokia Bell Labs for the discussion and collaboration. 

\bibliographystyle{ieeetr}
\bibliography{References}
\end{document}